\documentclass[graybox]{svmult}

\usepackage{mathptmx}       
\usepackage{helvet}         
\usepackage{courier}        
\usepackage{type1cm}        
\usepackage{makeidx}         
\usepackage{graphicx}        
\usepackage{multicol}        
\usepackage[bottom]{footmisc}
\usepackage{amsfonts, amsmath,amssymb}
\usepackage{graphicx}
\graphicspath{{Figures/}}
\newcommand{\one}{\mbox{$1 \hspace{-1.0mm}  {\bf l}$}}

\makeindex             

\begin{document}

\title*{ Non-Hermitian Quantum  Annealing and Superradiance}
\author{Alexander I. Nesterov, Gennady P. Berman, Ferm\'in Aceves 
de la Cruz, and Juan Carlos Beas Zepeda}
\authorrunning{A. I. Nesterov, G. P.  Berman, F. Aceves 
de la Cruz, and J. C. Beas Zepeda} 
\institute{Alexander I. Nesterov \at Departamento de F{\'\i}sica, CUCEI, 
Universidad de Guadalajara,
Av. Revoluci\'on 1500, Guadalajara, CP 44420, Jalisco, M\'exico 
\email{nesterov@cencar.udg.mx}
\and Gennady P.  Berman \at Theoretical Division, Los Alamos National 
Laboratory, and the New Mexico Consortium, Los Alamos, NM 87544, USA
\email{gpb@lanl.gov}
\and Ferm\'in Aceves 
de la Cruz \at Departamento de F{\'\i}sica, CUCEI, 
Universidad de Guadalajara,
Av. Revoluci\'on 1500, Guadalajara, CP 44420, Jalisco, M\'exico 
\email{fermin771009@gmail.com},
\and Juan Carlos Beas Zepeda\at Departamento de F{\'\i}sica, CUCEI, 
Universidad de Guadalajara,
Av. Revoluci\'on 1500, Guadalajara, CP 44420, Jalisco, M\'exico 
\email{juancarlosbeas@gmail.com}
}
%
%
\maketitle

\abstract{We consider  the non-Hermitian quantum annealing for the 
one-dimensional Ising spin chain, and for a large number of qubits. We 
show that the annealing time is significanlty reduced for the 
non-Hermitian algorithm in comparison with the Hermitian one. We also 
demonstrate the relation of the non-Hermitian quantum annealing with the 
superradiance transition in this system.
}

\section{Introduction}
\label{sec:1}

It is generally recognized that quantum annealing (QA) algorithms can be useful for solving many hard problems related to optimization of complex networks, finding the global minimum of multi-valued functions, and cost minimization \cite{KN,FGGLL,SSMO,DC,SMTC,OMNH,TSTR}. Opposite to classical annealing algorithm, the QA operates at zero temperature. Then, one can reformulate the optimization problem in terms of finding the ground state of the $N$-qubis system governed by the effective Hamiltonian.   

The idea of the  QA algorithm can be formulated as follows. Consider the  time-dependent Hamiltonian, $\mathcal H(t) =\mathcal H_0 +  \Gamma(t)\mathcal H_1$. Here $\mathcal H_0$ is the Hamiltonian to be optimized, $\mathcal H_1$ is an 
auxiliary (``initial") Hamiltonian, and it is assumed that $[\mathcal H_0,\mathcal H_1] \neq 0$.  
During the quantum annealing,  the external time-dependent field, $\Gamma (t)$, decreases from a large enough value to zero.  Then, the ground state of ${\rm {\mathcal H}}_{1} $ at $t=0$ (which is assumed to be known), can be considered as the initial (ground) state for the whole system. If $\Gamma(t)$ decreases sufficiently slowly, the adiabatic theorem guarantees finding the ground state of the main Hamiltonian, 
$\mathcal H_0$, at the end of computation.

 One of the main challenges is to accelerate the speed of QA algorithms, so that the annealing time grows not exponentially, but polynomially with the size of the problem \cite{SST,SNS,AM,SUD,JKKM,YKS}. In  \cite{BN}, we proposed the non-Hermitian quantum annealing (NQA) algorithm, which leads to a significant reduction of the annealing time. In NQA it is assumed that the non-Hermiticity vanishes at the end of the time-evolution. So, after the annealing is finished, the system is governed by the Hermitian Hamiltonian.
 
 Recently, we have applied the NQA algorithm to Grover's problem of finding a marked item in an unsorted database, and to study the transition to the ground state in a 1-dimensional ferromagnetic (and anti-ferromagnetic) Ising chain \cite{NABG,NABJBG,NBBB}. Analytical and numerical results  demonstrate that, even for a moderate value of the decay parameter, the NQA has a complexity of order $\ln N$, where $N$ is the number of qubits. This encouraging result is important in using classical computers in combination with quantum algorithms for  fast solutions of NP-complete problems. 

One of the open problems in the NQA is a dependence of annealing time on the value of the decay parameter. The answer is not obvious due to interference of the NQA with the superradiance transition (ST). The ST is associated with a significant enhancement of the spontaneous radiation due to quantum coherent effect as it first was shown by Dicke in 1954 \cite{Dicke}. Later it was demonstrated that the ST occurs in many quantum optical systems, nuclear systems (heavy nuclei decay), nano- and bio-systems \cite{VZA,VZ1,VZ2,Zel1,VZ3,Ber3,NABG1,BNLS,SL}. In these systems, the ST occurs when the discrete energy states of the system (associated with the quantum register) interact with the continuum spectra (associated with the sinks). An adequate approach for describing the dynamics of such systems can be based on an effective non-Hermitian Hamiltonian \cite{VZA,VZ1,VZ2,Zel1,VZ3,Ber3}. 

In this paper, we describe in details the NQA for finding the ground state of the ferromagnetic 1D chain of $1/2$ spins, and establish the relation of this problem with the ST in this system. The paper is organized as follows. In Sec II, we describe the model and the conditions of the applicability of the NQA. In Sec. III, we describe the dynamics of the NQA,  present the results of the numerical simulations, and discuss the relation between the NQA and the ST. In Conclusion, we discuss our results.

\section{ Non-Hermitian Quantum Annealing}

The generic adiabatic quantum optimization problem, based on the QA 
algorithm, can be formulated as follows \cite{BN}. Let $\mathcal H_0$ be 
the Hermitian Hamiltonian whose ground state is to be found. Consider the 
non-Hermitian time-dependent Hamiltonian:
\begin{equation}\label{NQA4}
 {\mathcal {\tilde H}_\tau(t)} = \mathcal H_0 + \mathcal{ \tilde H}_1(t/\tau),
\end{equation}
where $[\mathcal H_0,\mathcal{\tilde  H}_1] \neq 0$. The evolution of the system is determined by the Schr\"odinger equation:
\begin{align}\label{Sch1}
i\frac{\partial }{\partial t}|\psi(t)\rangle = {\mathcal {\tilde H}_\tau(t)} |\psi(t)\rangle.
\end{align}
The initial conditions are imposed as follows: $|\psi(0)\rangle=| \psi_g\rangle$, where $|\psi_g\rangle$ is the ground state of the auxiliary non-Hermitian Hamiltonian: $\mathcal{\tilde H}_1(0)|\psi_g\rangle= E_g|\psi_g\rangle$. At the end of the evolution, the total Hamiltonian, ${\mathcal {\tilde H}_\tau(\tau)} = \mathcal H_0 $, and the adiabatic 
theorem guarantees that the final state will be the ground state of $\mathcal H_0$, 
if the evolution was slow enough.

We denote by $|\psi_n(t)\rangle$ and $\langle\tilde\psi_n(t)|$ the right and the 
left instantaneous eigenvectors of the total Hamiltonian:
$   {\mathcal {\tilde H}_\tau(t)}|\psi_n (t)\rangle = E_n(t)|\psi_n\rangle$, 
$\langle\tilde\psi_n (t)| {\mathcal {\tilde H}_\tau(t)}=
    \langle\tilde\psi_n(t)|E_n(t)$.
We assume that these eigenvectors form a bi-orthonormal basis,
$\langle\tilde\psi_m|\psi_{n}\rangle = \delta_{mn}$ \cite{MF}. For the non-Hermitian quantum optimization problem, governed by the 
Hamiltonian (\ref{NQA4}), the validity of the adiabatic approximation requires,
\begin{equation}\label{NQA3}
  \tau \gg \frac{\max |\langle \tilde \psi_e(t) |\mathcal {\dot {\tilde 
  H}_\tau}(t)|\psi_g(t)\rangle|}{\min |E_e(t)- E_g(t)|^2},
\end{equation}
where ``dot'' denotes the derivative with respect to the dimensionless 
time, $s=t/\tau$, and $E_e$ is the energy of the first excited state, 
$|\psi_e\rangle$. This restriction is violated near the ground state 
degeneracy, where complex energy levels cross. The degeneracy is 
known as the exceptional points (EP), and it is characterized by a coalescence of 
eigenvalues and their corresponding eigenvectors, as well 
\cite{KT,H0,H1,B,IR,SKM,KMS,MKS1}. Thus, if  the gap, $\Delta E = 
\min |E_e - E_g|$, is small enough, the time required to pass from the initial state 
to the final state becomes very large, and the NQA loses its advantage over the
thermal annealing.

\subsection{Description of the model}

Consider the time-dependent Hamiltonian, 
 $ \mathcal {\tilde H}_\tau(t)= \sum_{k}\mathcal {\tilde H}_k(t)$, where
  \begin{align}\label{H1g}
 \mathcal {\tilde H}_k(t)  = -{\varepsilon_{0k}(t)} \one + {J} \left(
\begin{array}{cc}
              \tilde g(t)- \cos  \varphi_k & \sin  \varphi_k \\
             \sin  \varphi_k & -\tilde g(t)+ \cos  \varphi_k  \\
            \end{array}
          \right),
\end{align}
 $\varepsilon_{0k}(t)= J\cos \varphi_k + iJ\delta(t)$, $\tilde g(t) = g(t) + 
i\delta(t)$, $\varphi_k = {\pi( 2k-1)}/{N}$, and $k$ takes the following discrete 
values: $ k=\pm 1,2,\dots, \pm{N}/{2}$,  $N/2$ being an even integer. Further, we asuume linear dependence of the function $\tilde g(t)$ on time:
\begin{align}
\tilde g(t) = \left\{
\begin{array}{l}
(g +i\delta) (1- t/\tau), \quad 0 \leq t \leq \tau, \\
0 , \quad t > \tau,
\end{array}
\right.
\end{align}
where $g$ and $\delta$ are real parameters. The Hamiltonian, $ \mathcal {\tilde H}_\tau(t)$, describes  the NQA for  the 1D Ising model in a transverse magnetic field \cite{NBBB}, in the momentum space. The external magnetic field is associated with the parameter, $g$, and the rate of decay into continuum (sink) is described by the parameter, $\delta$.

Let $g \gg 1$, then the initial ground state of the 
total Hamiltonian is determined by the ground state of the auxiliary Hamiltonian, ${\mathcal {\tilde H}}_1(0) = J\sum_k ( -i\delta +( g+ i\delta)\sigma_z)$.  At the end of the NQA, one obtains, $\mathcal {\tilde H}_\tau(\tau)= {\mathcal H}_0 $, where 
\begin{align}
	{\mathcal H}_0 = J\sum_k (-\cos\varphi_k \one +\sin\varphi_k\sigma_x -\cos\varphi_k\sigma_z).
\end{align}
If the quench is slow enough, the adiabatic theorem guarantees the approach of
the ground state of the main Hamiltonian, $\mathcal H_0$, at the 
end of the computation.

Resolving the eigenvalue problem for the Hamiltonian, $\mathcal H_k(t)$,
we obtain
\begin{align}\label{E2a}
 &|u_{+}(k,t)\rangle = \left(\begin{array}{c}
                 \cos\frac{\theta_k(t)}{2} \\
                  \sin\frac{\theta_k (t)}{2}
                  \end{array}\right), \,
\langle \widetilde u_{+}(k)| = \big(\cos\frac{\theta_k (t)}{2} ,
\sin\frac{\theta_k (t)}{2} \big),\\
&|u_{-}(k,t)\rangle = \left(\begin{array}{c}
               -\sin\frac{\theta_k (t)}{2}\\
                 \cos\frac{\theta_k (t)}{2}
                  \end{array}\right), \,
\langle \widetilde u_{-}(k,t)| = \big(-\sin\frac{\theta_k (t)}{2}, 
\cos\frac{\theta_k (t)}{2}
 \big),
 \end{align}
 where
 \begin{align}\label{Th}
\cos\theta_k(t) = \frac{\tilde g(t) -  \cos \varphi_k }{\sqrt{\tilde g^2(t) -  2\tilde 
g(t)\cos  \varphi_k +1}}, \\
\sin\theta_k(t) = \frac{ \sin  \varphi_k}{\sqrt{\tilde g^2(t) -  2\tilde g(t)\cos  
\varphi_k +1}},
\end{align}
$\theta_k$ being a complex angle. We denote by $ |u_{\pm}(k,t)\rangle$ 
($\langle \widetilde u_{\pm}(k,t)| $) the right (left) instantaneous eigenvectors of the 
Hamiltonian $\mathcal H_k(t)$, and by $\varepsilon_{\pm}(k,t)= -\varepsilon_{0k}(t) \pm 
\varepsilon_k(t)$ the corresponding (complex) eigenenergies, 
\begin{align}
	\varepsilon_{0k} (t)&= J\cos \varphi_k+ iJ\delta(1-t/\tau), \\
	\varepsilon_k (t)&= J\sqrt{\tilde g^2(t) -  2\tilde g(t)\cos \varphi_k +1}.
  \label{E2b}
\end{align}
\begin{figure}[b]
\scalebox{0.25}{\includegraphics{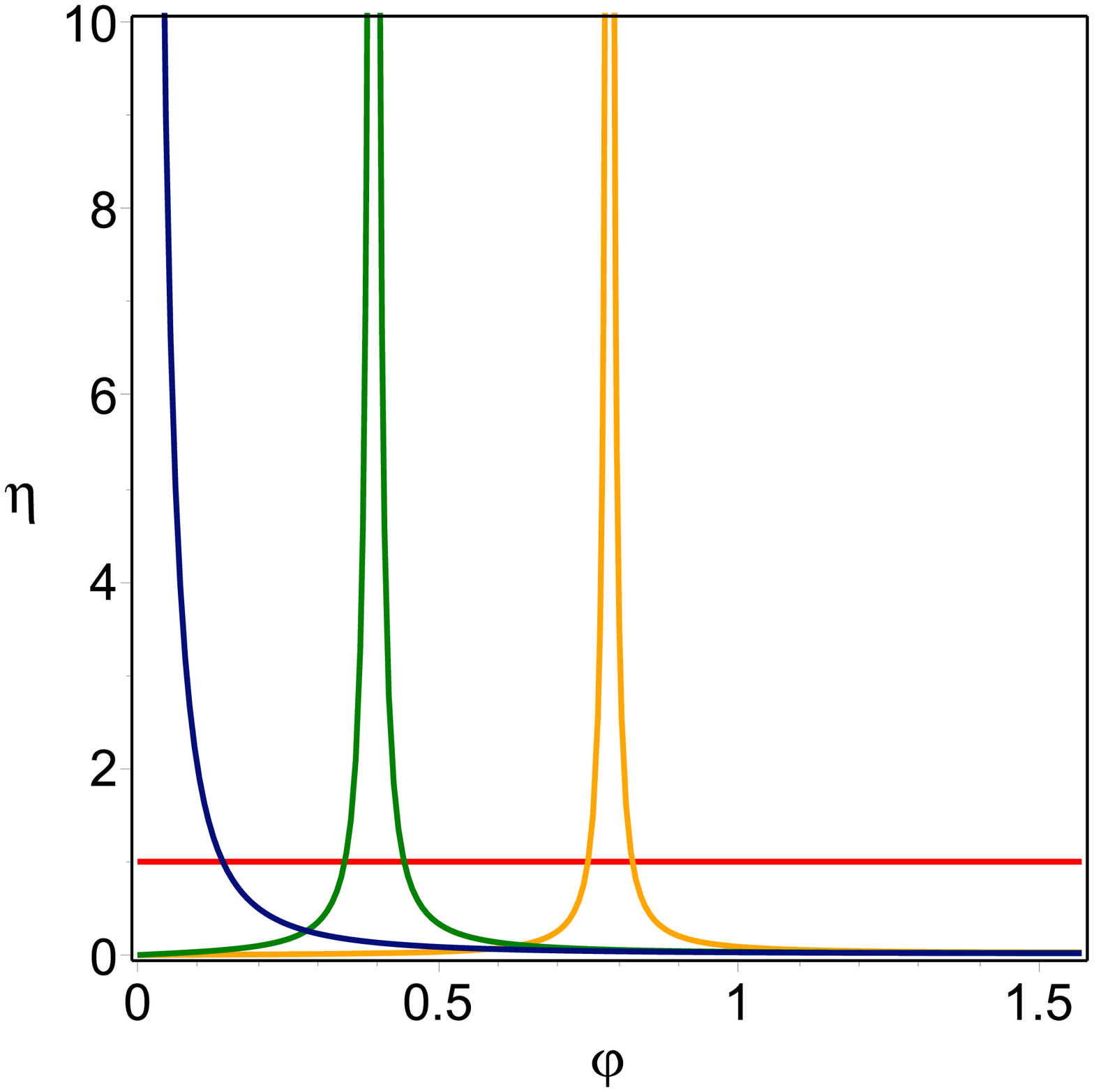}}
(a)
\scalebox{0.325}{\includegraphics{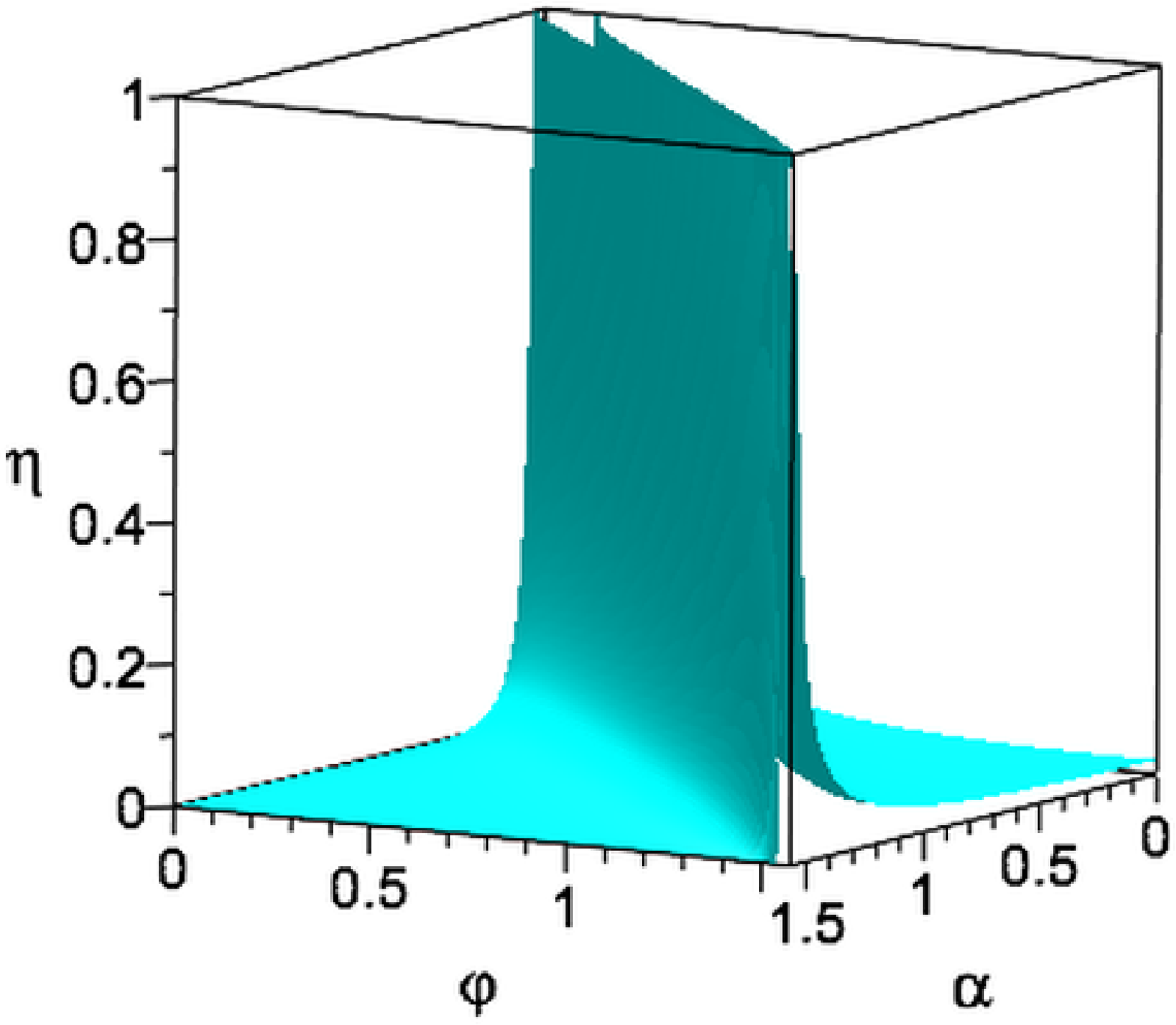}}
(b)
\caption{(a) The dependence of the function $\eta$ on  $\varphi$: $\alpha 
=0$ (blue), $\alpha =\pi/8$ (green), $\alpha =\pi/4$ (orange). Red line 
corresponds to $\eta=1$. (b) The dependence of the function $\eta$ on 
$\alpha$ and $\varphi$. Parameters: $J=0.5$, $g=10$, $\tau = 500$.}  
\label{ADT1a}
\end{figure}

For a given value of $k$, the complex energy levels of the effective non-Hermitian Hamiltonian, $\mathcal H_k(t)$, cross at the EP, which is  defined by the condition: $\tilde g(t_c) =e^{\pm \varphi_k}$.  As results, the energy gap vanishes at the  EP. The computation yields $ {\varphi_k}_c = \tan^{-1} (\delta/g)$. The system reaches the EP at the moment of time
\begin{align}
	 \frac{t_c}{\tau} =1- \frac{1}{\sqrt{g^2 + \delta^2}} .
\end{align}
Note, that the difference between the Hermitian QA and non-Hermitian QA is that, while in the first case the minimal gap occurs for long wavelength modes ($k \sim 1 $), in the second case the minimal gap shifts to short wavelength modes ($k \sim (N/2\pi) \tan^{-1} (\delta/g)$). In particular, for $\delta \gg g $ we obtain $k \sim N/4$.

The requirement of the adiabatic theorem (\ref{NQA3}) can be rewritten in the equivalent form as,
\begin{align}\label{QA3a}
\max\bigg | \frac{d\theta_k}{dt} \bigg| \ll \min 2|\varepsilon_k|.
    \end{align}
 We find
\begin{align}
	\max\bigg | \frac{d\theta_k}{dt} \bigg| = & \frac{g \cos\alpha |\sin\varphi_k|}{\tau |\sin(\alpha -\varphi_k)\sin(\alpha +\varphi_k)|}, \\
	 \min |\varepsilon_k|=&\frac{ J}{\cos\alpha}\sqrt{|\sin(\alpha -\varphi_k)\sin(\alpha +\varphi_k)|},
\label{Theta1}
\end{align}
where $\alpha = \tan^{-1}(\delta/g)$. Employing these results, one can recast Eq. (\ref{QA3a}) as,
\begin{align}
\eta_k = \frac{g \cos^2\alpha |\sin\varphi_k|}{2J\tau (|\sin(\alpha -\varphi_k)\sin(\alpha +\varphi_k)|)^{3/2}} \ll 1.
 \label{Ad}
\end{align}
Thus, for a given value of $k$, the condition of the adiabaticity  can be written as $\eta_k\ll 1$. This condition is violated in the vicinity of the EP, defined by the equation, $\alpha = \varphi_k$. In particular, for the Hermitian QA ($\alpha =0$), we obtain
\begin{align}
\eta_k = \frac{g }{2J\tau \sin^2 \varphi_k} \ll 1.
 \label{Ad}
\end{align}
Since, $\eta_k \leq \eta_1$, we find that the adiabatic approximation is valid for all modes, if
\begin{align}
\frac{g }{2J\tau \sin^2( {\pi }/{N} )} \ll 1.
 \label{Ad1}
\end{align}
For $N\gg 1$, we obtain the following estimate of the annealing time,
\begin{align}
\tau \gg \frac{g N^2}{2 J\pi^2}.
\label{omega2}
\end{align}

In the thermodynamic limit ($N\gg1$), one can consider the variable $\varphi_k$ as a continuous variable, $\varphi$. In Fig. \ref{ADT1a}, the adiabatic parameter, $\eta$, as a function of the angles, $\varphi$ and $\delta$, is presented. One can see the regions of the adiabatic conditions, when $\eta\ll 1$.

\section{ Quench dynamics}

Since the Hamiltonian of the system is presented by the sum of the independent terms, the total wave functions, $|\psi\rangle $ and $\langle\tilde\psi |$  ,  can be written as the product: $|\psi\rangle = \prod_{k}|\psi_k\rangle$ and  $\langle\tilde\psi |= \prod_{k}\langle\tilde\psi_k|$. The wave functions,  $|\psi_k\rangle$ and  $\langle\tilde \psi_k |$, satisfy the Schr\"odinger equation and its adjoint equation:
 \begin{eqnarray}\label{Eqh1}
i\frac{\partial }{\partial t}|\psi_k\rangle&= \mathcal {\tilde H}_k(t)|\psi_k\rangle , \quad
-i\frac{\partial }{\partial t}\langle\tilde \psi_k |& =
\langle\tilde \psi_k |\mathcal {\tilde H}_k(t)\label{NS2}.
\end{eqnarray}
Presenting $|\psi_k(t)\rangle$ as, 
\begin{align}\label{S1}
|\psi_k(t)\rangle = (u_k(t)|0\rangle + v_k(t)|1\rangle) e^{i\int \varepsilon_0(t)dt},
\end{align}
and inserting expression (\ref{S1}) into Eq. (\ref{Eqh1}), we obtain
\begin{align}\label{IS2a}
i\dot  u_k &= J\big(-(\tilde  g - \cos \varphi_k)\,u_k + \sin \varphi_k\, v_k\big), \\
i\dot  v_k &= J\big(\sin \varphi_k\, u_k +(\tilde  g - \cos \varphi_k)\,v_k \big ).
\label{IS2b}
\end{align}
The solution can be written in terms of  the parabolic cylinder functions, $D_{-i\nu_k}(\pm z)$. (For details see \cite{NABJBG}.)

In the adiabatic basis formed by the instantaneous eigenvectors of the Hamiltonian $\mathcal H_k(t)$,  the  wavefunction, $|\psi_k(t) \rangle $, can be written as,
\begin{eqnarray}\label{Eq7}
 |\psi_k(t) \rangle =( \alpha_k(t) |u_{-}(k,t)\rangle + \beta_k(t) |u_{+}(k,t)\rangle)e^{i\int \varepsilon_{0k}(t)dt}.
\end{eqnarray}
From Eqs. (\ref{E2a}) and (\ref{S1}) it follows
\begin{eqnarray}\label{Eq7a}
\alpha_k(t) =u_k(t)\cos\frac{\theta_k(t)}{2}- v_k(t)\sin\frac{\theta_k(t)}{2}, \\
\beta_k(t) =v_k(t)\cos\frac{\theta_k(t)}{2}+ u_k(t)\sin\frac{\theta_k(t)}{2}.
\end{eqnarray}

Then one can show that
\begin{align}
|\Psi_k(t)\rangle = \left (\begin{array}{c}
  \beta_k(t) \\
   \alpha_k(t) 
 \end{array}
\right ).
\end{align}
satisfies the Schr\"odinger equation
 \begin{eqnarray}\label{Eq3A}
i\frac{\partial }{\partial t}|\Psi_k\rangle&= {\mathcal { H}}_k(t)|\Psi_k\rangle ,
\end{eqnarray}
with the new Hamiltonian
\begin{align}
\label{Eq8}
{\mathcal {H}}_k(t) = \left(
\begin{array}{cc}
          \varepsilon_k & i\dot \theta_k/2\\
           -i\dot \theta_k/2& - \varepsilon_k  \\
            \end{array}
          \right).
\end{align}
Using the relation, $\tan \theta_k = \sin\varphi_k/(\tilde g(t) \cos\varphi_k)$, we obtain
\begin{align}
\frac{d\theta_k}{dt} = -\frac{\dot {\tilde g} (t) \sin^2\theta_k(t)}{\sin\varphi_k}.
\label{Theta}
\end{align}
  
Assume that the evolution begins from the ground state. This implies 
$\alpha_k(0)=1$ and $\beta_k(0)=0$. Then, the solution of the  Eq. (\ref{Eq3A}) can be written as \cite{NABJBG},
\begin{align}\label{eq5g}
\alpha_k(t) =& C_k \bigg( D_{-i\nu_k}(z_k(t))\sin \frac{ \varphi_k}{2} 
 + \sqrt{i\nu_k }  D_{-i\nu_k-1}(z_k(t))\cos\frac{ \varphi_k}{2}    \bigg) 
,\\
\beta_k(t) =& C_k \bigg( D_{-i\nu_k}(z_k(t))\cos\frac{ \varphi_k}{2} 
- \sqrt{i\nu_k }  D_{-i\nu_k-1}(z_k(t))\sin\frac{ \varphi_k}{2}    \bigg),
\label{eq5e}
 \end{align}
where
\begin{align}\label{Eq4b}
z_k(t) =& e^{i\pi/4}\sqrt{\frac{2\tau J}{g+ i\delta}}\big((g+ i\delta)(1- t/\tau)-\cos \varphi_k\big), \\
\nu_k=& \frac{\tau J\sin^2 \varphi_k}{2(g+ i\delta)}.
\label{Eq4c}
\end{align}
At the end of evolution at $t=\tau$, when $\tilde g=0$, we obtain\begin{align}\label{E3}
      |\psi_k(\tau) \rangle= \alpha_k(\tau) |u_{-}(k,\tau)\rangle + 
      \beta_k(\tau) |u_{+}(k,\tau)\rangle,
\end{align}
where
\begin{align}\label{Eq1}
|u_{+}(k,\tau)\rangle =\left (\begin{array}{c}
                \sin\frac{ \varphi_k}{2}  \\
             \cos \frac{ \varphi_k}{2}\\
            \end{array}
          \right),  \quad 
|u_{-}(k,\tau)\rangle = \left (\begin{array}{c}
               - \cos\frac{ \varphi_k}{2}  \\
             \sin \frac{ \varphi_k}{2}\\
            \end{array}
          \right).
\end{align}

Since for the non-Hermitian systems the norm of the wave function is not conserved, we define the partial survival probability as  \cite{NABJBG,NBBB},
\begin{align}\label{Eq2}
P^{gs}_k(t) = \frac{|\alpha_k(t)|^2}{|\alpha_k(t)|^2 + 
|\beta_k(t)|^2}.
\end{align}
 The probability of the whole system to stay in the ground state at the end of the evolution is the 
product:
\begin{align} \label{Eq4}
P_{gs} = \prod_{k>0} P^{gs}_k(\tau).
\end{align}
  
For long wavelength modes with $|\varphi_k|\ll \pi/4$, using the asymptotic formulas for the Weber functions with the large value of its argument, one has \cite{NABJBG,NBBB}
\begin{align}\label{PA1}
P_k(\tau) \approx\frac{1}{1+ \displaystyle\frac{|\Gamma(1+i\nu_k )|^2 }{2\pi |\nu_k| }\,e^{-\pi\Re\nu_k- \Re z^2_k(\tau)}}.
\end{align}

In the long wavelength approximation, one can take into account only the first mode, $\varphi_1= \pi/N$, and   estimate $ P_{gs}$ as,
\begin{align}\label{PGS1}
P_{gs} \approx\frac{1}{1+ \displaystyle\frac{\tau|\Gamma(i\nu )|^2 }{2\pi \tau_0}\,e^{-\pi\kappa}},
\end{align}
where  $\tau_0 = 2gN^2/(\pi^2J)$, $\nu =\cos\alpha e^{-i\alpha} \tau/\tau_0$ and $\kappa = (\tau/\tau_0) \cos^2\alpha + (\tau J/\pi g )\sin(2\alpha)$.
 For the Hermitian QA this yields the Landau-Zener 
formula \cite{LL,ZC}
\begin{align}\label{P}
P_{gs} =1-e^{-2\pi\tau/\tau_0}.
\end{align}
For $\tau \gtrsim \tau_0 $ we obtain $P_{gs} \approx 1$. Thus, 
the computational time for the Hermitian QA should be of order $N^2$.

When $\tau \ll \tau_0$, one can approximate the Gamma function as, $|\Gamma(i\nu )| \approx 1/|\nu |$, and rewrite  Eq. (\ref{PGS1})  as,
\begin{align}\label{PGS1a}
P_{gs} =\frac{1}{1+ \displaystyle\frac{\tau_0 }{2\pi \tau}\,e^{-\pi\kappa_0}} .
\end{align}
where $\kappa_0 =  (\tau J/\pi g )\sin(2\alpha)$. Assuming $e^{-\pi\kappa_0} \ll {2\pi \tau}/\tau_0$, we obtain
\begin{align}\label{PGS1b}
P_{gs} \approx 1- \displaystyle\frac{\tau_0 }{2\pi \tau}\,e^{-\pi\kappa} .
\end{align}
 If the condition, 
\begin{align}\label{AT1}
\frac { \tau J}{g} \sin(2\alpha)- \ln \frac{\tau_0}{2\pi\tau} \gg 1,
\end{align}
is satisfied, one has $P_{gs} \approx 1$. From (\ref{AT1}), we obtain the following rough estimate of the computational time for NQA: 
\begin{align}\label{AT1a}
\tau \approx  \frac{2g \ln ( N/\pi) }{J\sin (2\alpha)} =  \frac{(g^2 + \delta^2 )\ln ( N/\pi) }{\delta J} .
\end{align}
Thus, while the Hermitian QA has complexity of order $N^2$,
 the NQA has complexity of order $\ln N$, which is much better.
\begin{figure}[tbh]
  \sidecaption
  \scalebox{0.165}{\includegraphics{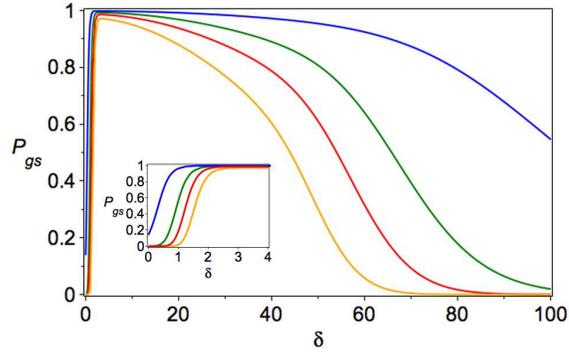}}
  \caption{ The probability to stay in the ground state, $P_{gs}$, as 
  a function of the decay parameter $\delta$:  $N=64\, (\rm blue)$, $256 \,
  (\rm green)$, $512\, (\rm red)$, $1024\, (\rm orange)$. Parameters: 
  $J=0.5$, $g=10$, $\tau = 500$. }
  \label{Pgs_3}
  \end{figure}
As one can see, there exists the optimal choice of the parameter $\alpha$, namely, $\alpha = \pi/4$ (or $\delta = g$), that minimizes the annealing time yielding $\tau\approx (g/ J)\ln ( N/\pi)$. For $\delta \ll g$ the annealing time can be estimated as, $\tau \approx (g^2/(\delta J))\ln ( N/\pi)$, and for $\delta \gg g$ we obtain, $\tau \approx (\delta/ J)\ln ( N/\pi)$.

In Figs. \ref{Pgs_3},  the results of numerical simulation are 
demonstrated for $N=64$, $256$, $512$, $1024$ qubits ($g=10$, $\tau = 500$). As one can see, for $\delta \sim g$, the probability to stay in the ground state at the end of evolution is: $P_{gs}\approx 1$. However, for  $\delta \ll g$ and $\delta \gg g$, the probability  to stay in the ground state significantly decreases.

\begin{figure}[b]
\sidecaption
\scalebox{0.25}{\includegraphics{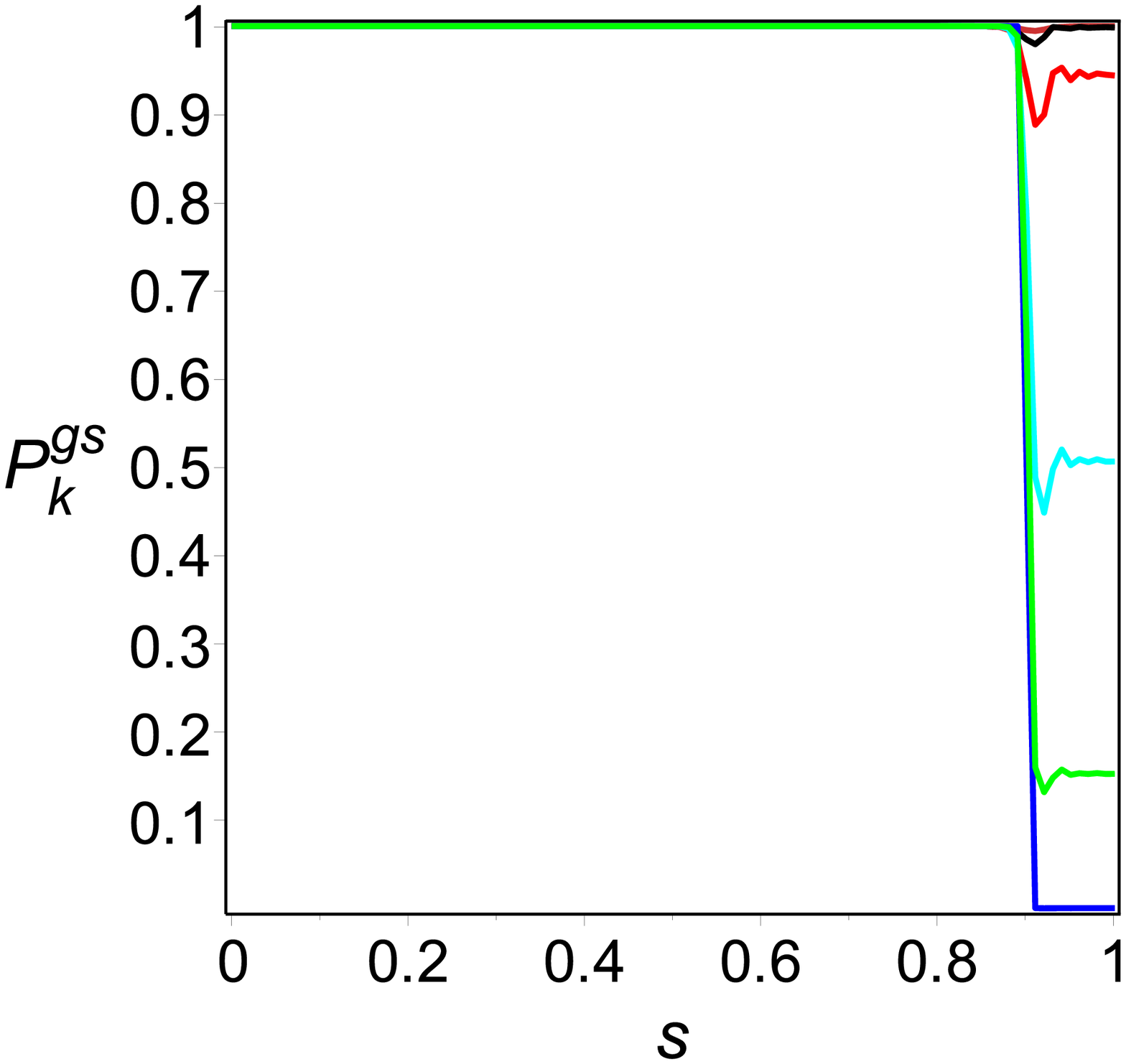}}
\caption{The probability, $P^{gs}_k$, to stay in the 
ground state as a function of the scaled time, $s =t/\tau $, for the 
Hermitian QA. Blue curve ($k=1$), green curve ($k=8$),  cyan line ($k=16$), red curve ($k=32$), black curve ($k=48$), orange curve ($k=64$). Parameters: $\delta = 0$, $J=0.5$, $g=10$, $\tau = 500$, $N=1024$. }
\label{PQA_1}
\end{figure}
 \begin{figure}[b]
\scalebox{0.295}{\includegraphics{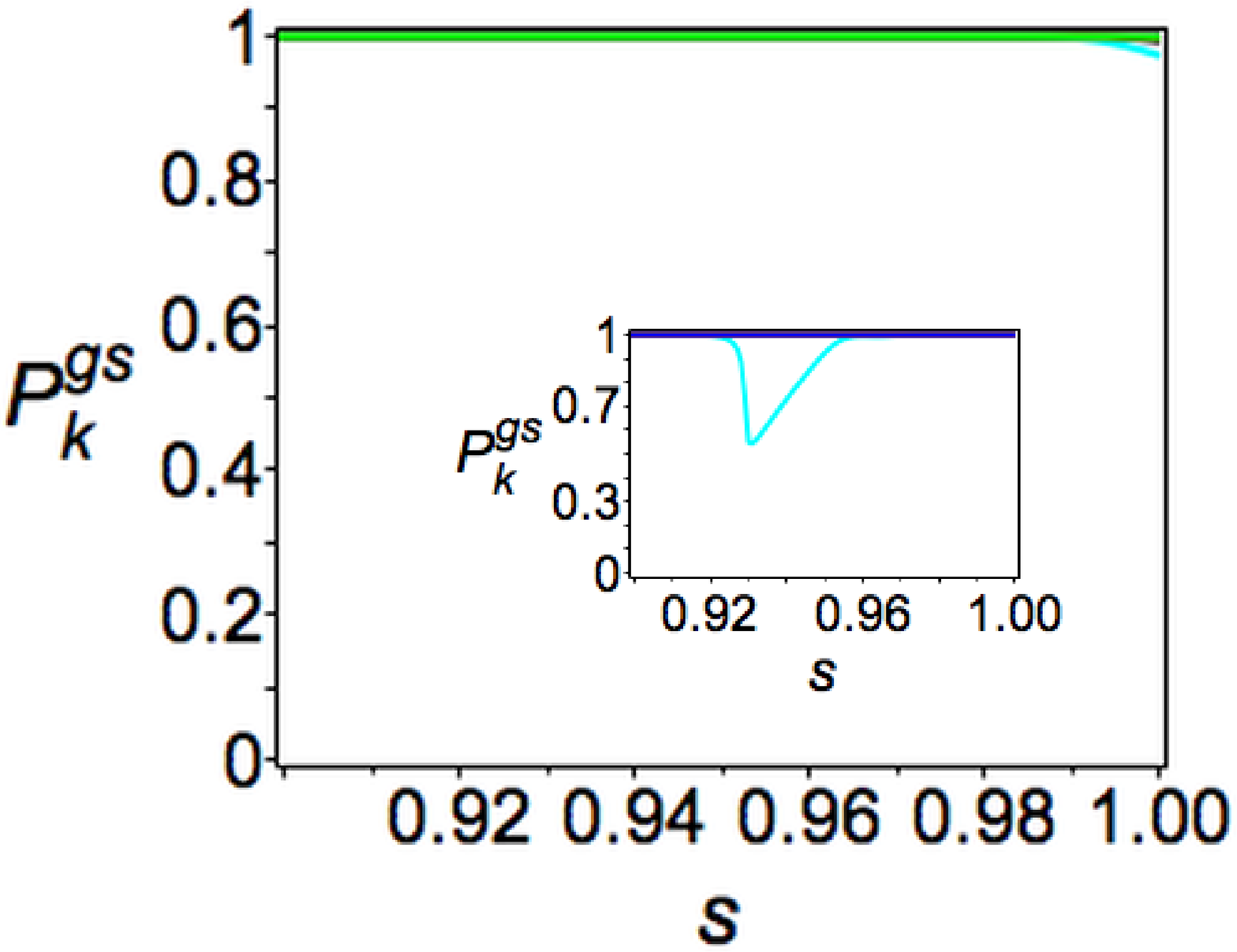}}
(a)
\scalebox{0.295}{\includegraphics{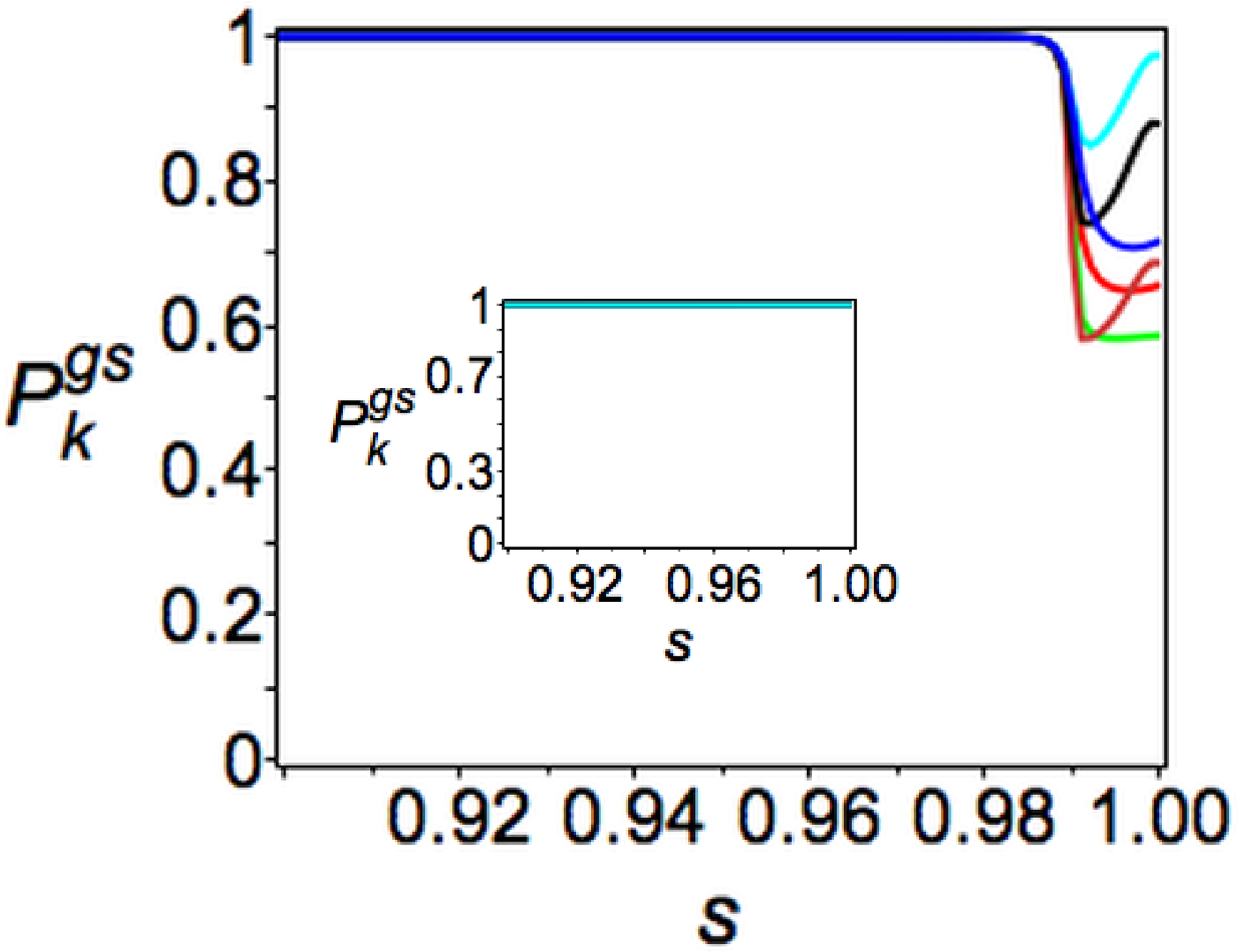}}
(b)
\caption{NQA: the probability, $P^{gs}_k$, to stay in the 
ground state as a function of the scaled time, $s=t/\tau$.
 (a) Blue curve ($k=1$), red curve 
($k=8$), green curve ($k=16$), orange curve ($k=32$), black 
curve ($k=64$), cyan line ($k=128$). (b)Blue curve ($k=224$), red curve 
($k=230$), green curve ($k=236$), orange curve ($k=242$), black 
curve ($k=248$), cyan curve ($k=256$). Parameters: $\delta = 100$ , $J=0.5$, $g=10$, 
$\tau = 500$, $N=1024$ (inset: $\delta =10$).}
\label{P2b}
\end{figure}
\begin{figure}[b]
\scalebox{0.3}{\includegraphics{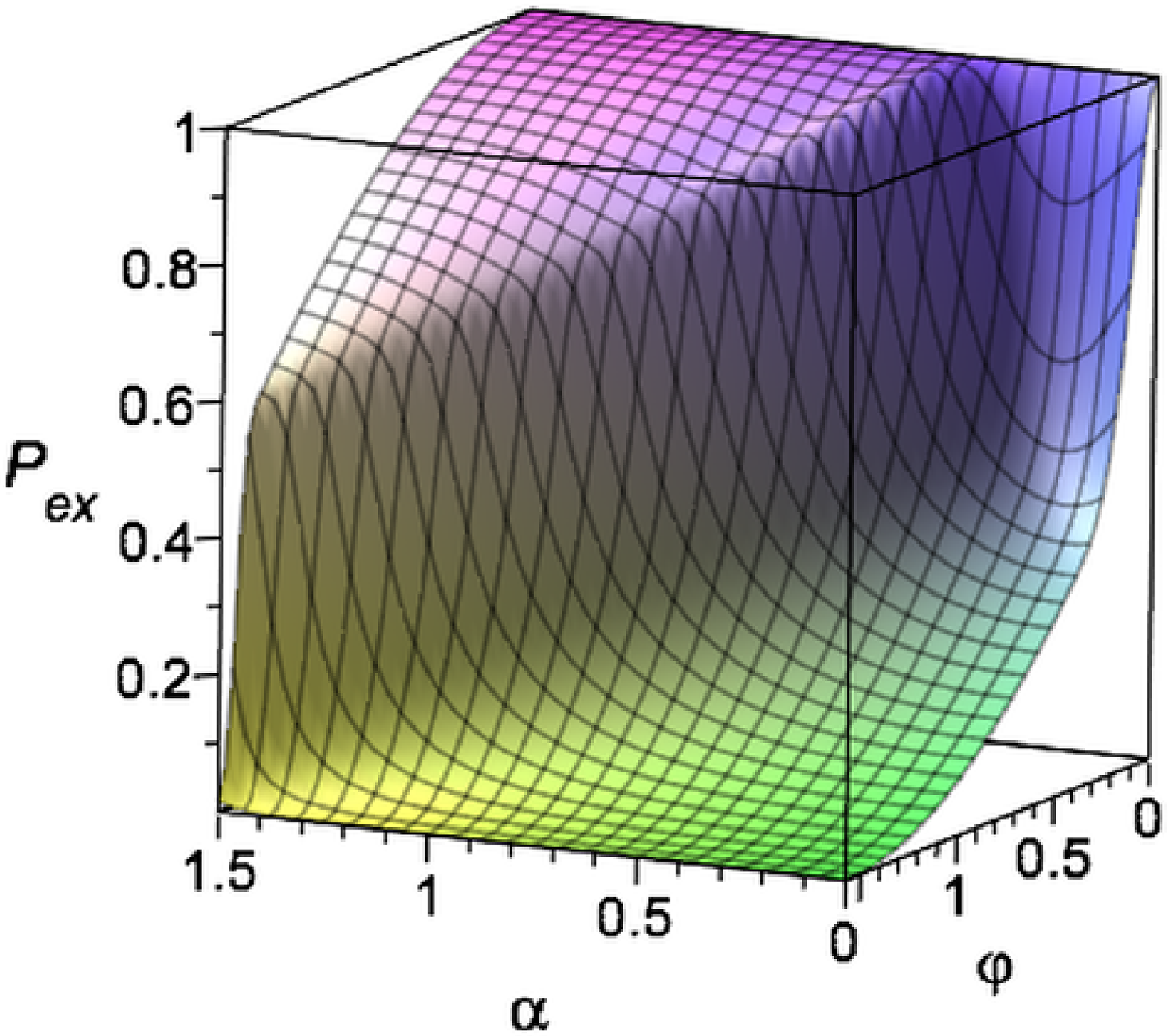}}
(a)
\scalebox{0.3}{\includegraphics{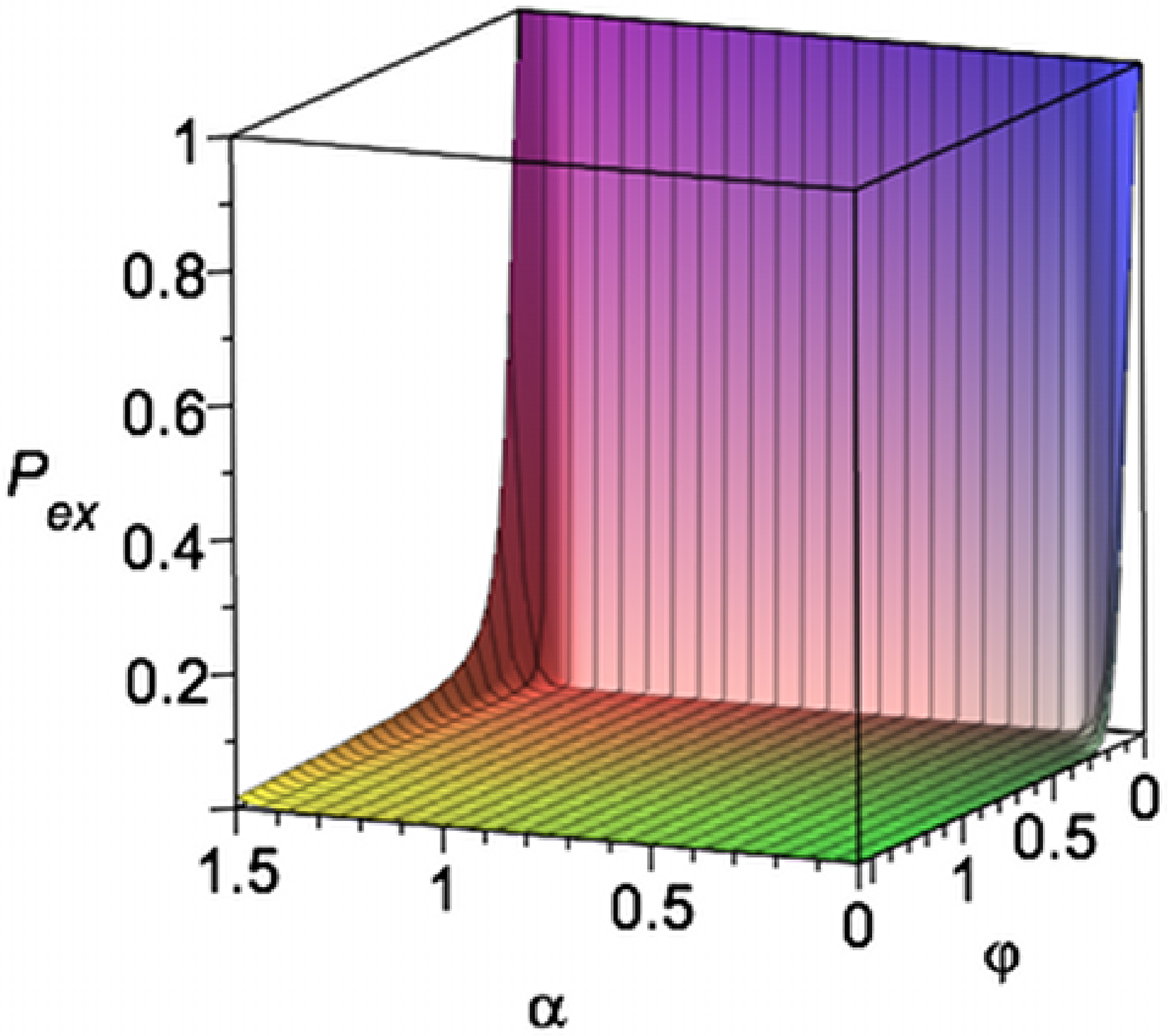}}
(b)
\caption{The probability, $P_{ex}$, of excited states as a function of $\alpha$ and $\varphi$. (a) $P_{ex}$ calculated at the critical point $t_c$. (b) $P_{ex}$ calculated at the end of evolution, at time $t_f = \tau$. Parameters:  $J=0.5$, $g=10$, 
$\tau = 500$.}
\label{P2c}
\end{figure}

Our theoretical predictions are confirmed by the results of our numerical 
calculations 
performed for $N=1024$ qubits, and presented in   Figs. \ref{PQA_1}, 
\ref{P2b}, and   \ref{P2c}. For the Hermitian QA, the long wave modes with 
$\varphi_k \ll \pi/4$ are excited (see Fig. \ref{PQA_1}). For $\delta \lesssim 
g$, one can observe that while short wavelength excitations are essential at 
the critical 
point, at the end of the evolution their contribution to the transition 
probability 
from the ground state to the first excited state is negligible. However, for 
$\delta \gg g$, the contribution of the shortwave excitations is important, 
and this results in violation of the adiabatic theorem. 
 
\subsection{Defects formation}

During the QA, the system does not stay always in the ground state at all 
times. At the critical point, the system becomes excited, and its final state is 
determined by the number of defects (kinks). To evaluate the efficiency of  the
QA  one can calculate the number of defects. Then, the computational time is 
the time required to achieve the number of defects below some acceptable 
value.
\begin{figure}[tbp]
\sidecaption
\scalebox{0.25}{\includegraphics{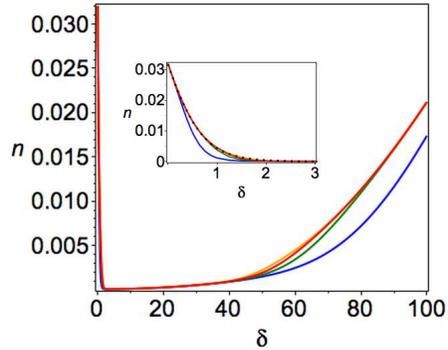}}
\caption{ Density of kinks as function of the decay 
parameter $\delta$:  $N=64\, (\rm blue)$, $256 \,
(\rm green)$, $512\, (\rm red)$, $1024\, (\rm orange)$. Inset: zoom of the 
figure (dotted black line presents the results of the Eq. (\ref{Eq10h})). 
Parameters: $J=0.5$, $g=10$, $\tau = 500$.}
\label{P2h}
\end{figure}
Following \cite{DJ}, we define the operator of the number of kinks as
\begin{align}
 \hat{\mathcal N} = \frac{1}{2}\sum^N_{n=1}\big ( 1- \sigma^z_n 
 \sigma^z_{n+1}\big ) .
\end{align}
The final number of kinks is equal to the number of quasiparticles
excited at the end of the evolution:
${\mathcal N} =  \langle \Psi_\tau |\hat{\mathcal N} | \Psi_\tau \rangle $. Using  Eq. (\ref{E3}), we obtain
\begin{align}\label{TL1}
\mathcal N = \sum_{k>0}( 1- P^{gs}_k(\tau)),
\end{align}
where $P^{gs}_k(\tau)$ is given by Eq. (\ref{Eq2}). In Fig. \ref{P2h}, the dependence of  the density of defects on decay parameter, $\delta$, and annealing time, $\tau$, is presented. As one can see, even moderate dissipation essentially decreases the number of defects in the system.

In the thermodynamic limit ($N\gg 1$) the sum in Eq. (\ref{TL1}) can be replaced by 
The integral, and we obtain for the density of kinks the following expression:
\begin{align}\label{QP1a}
n= \lim_{N\rightarrow \infty} \frac{\mathcal N}{N} = 
\frac{1}{\pi}\int^{\pi}_{0} d k( 1- P^{gs}_k(\tau)).
\end{align}
In the limit $\sqrt{2J\tau/|g + i\delta|} \gg 1$,
only the long wavelength modes  yield  the main contribution,  and one can use the Gaussian approximation to calculate the integral. By performing the integration, one obtains \cite{NABJBG}\begin{align}\label{Eq10h}
n =  n_0 e^{-2\delta\tau J/g^2}\Phi\Big( 1-e^{-2\delta\tau 
J/g^2},\frac{1}{2},1\Big),
\end{align}
where,
\begin{align}\label{n1}
 n_0 = \frac{1}{2\pi}\sqrt{\frac{g}{J\tau}},
\end{align}
denotes the density of kinks for the Hermitian LZ problem \cite{DJ}, and  
$\Phi(x,a,c)$ is the Lerch transcendent \cite{EMOT1}. The approximated 
formula (\ref{Eq10h}) is good enough for $\delta \lesssim g $ (see the inset 
in Fig. \ref{P2h}).

\section{Superradiance and NQA  }

\begin{figure}[b]
\sidecaption
\scalebox{0.3}{\includegraphics{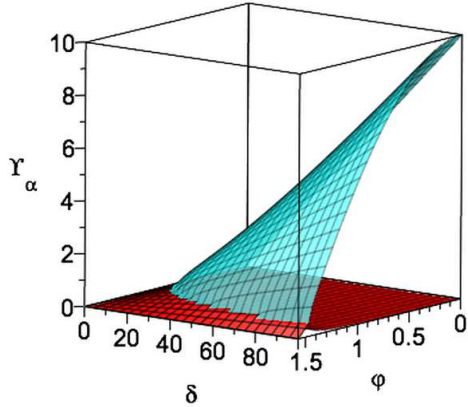}}
\caption{Width $\Upsilon_\alpha$ ($\alpha=1,2$) as the function of the decay 
rate $\delta$  and $\varphi$ in the thermodynamic limit, calculated at the 
exceptional point: $\Upsilon_1$ (red surface) and $\Upsilon_2$ (cyan 
surface). Choice of parameters: $J=0.5$, $g=10$, $\tau = 500$.}
\label{ADT2}
\end{figure}

\begin{figure}[b]
\scalebox{0.315}{\includegraphics{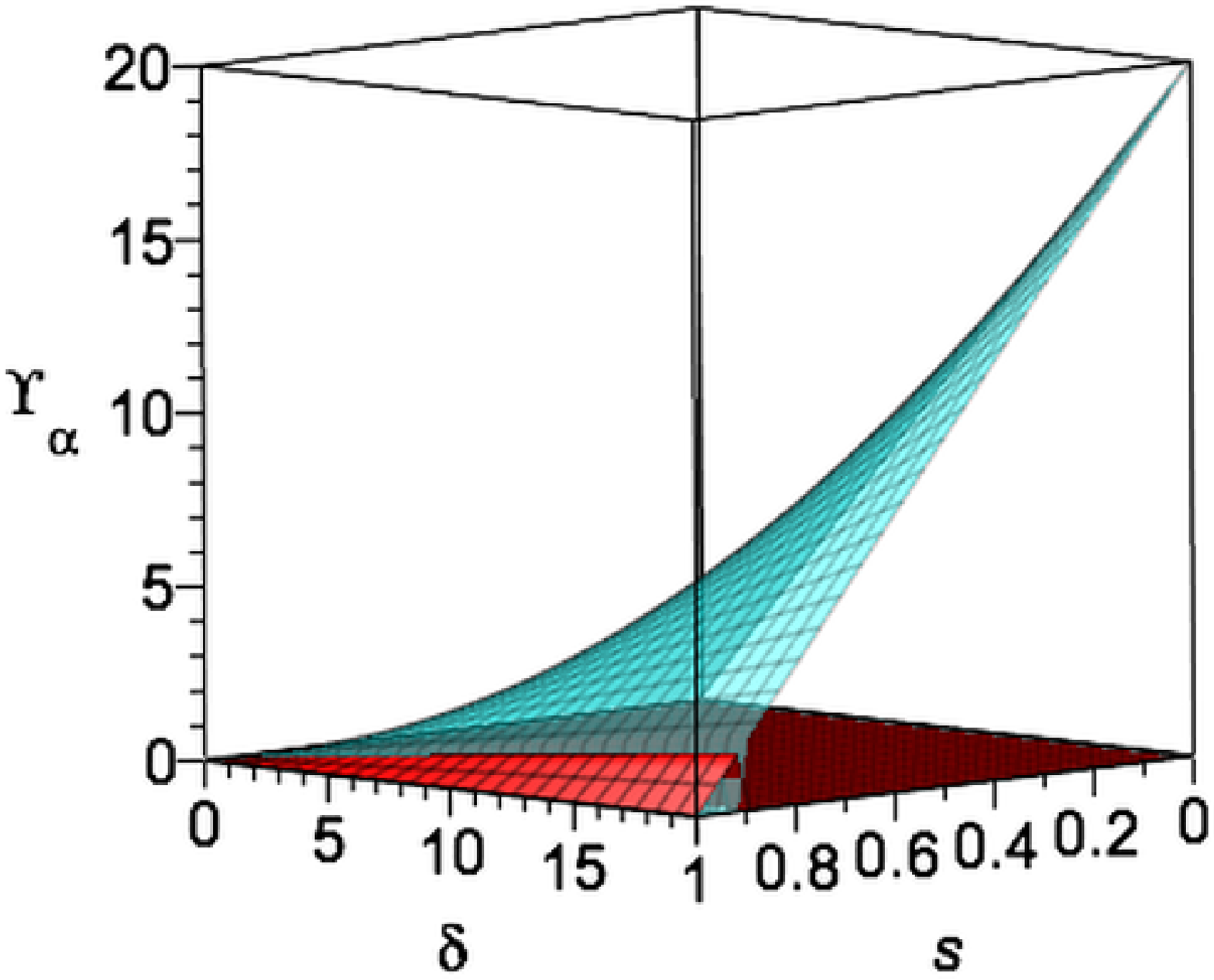}}
(a)
\scalebox{0.315}{\includegraphics{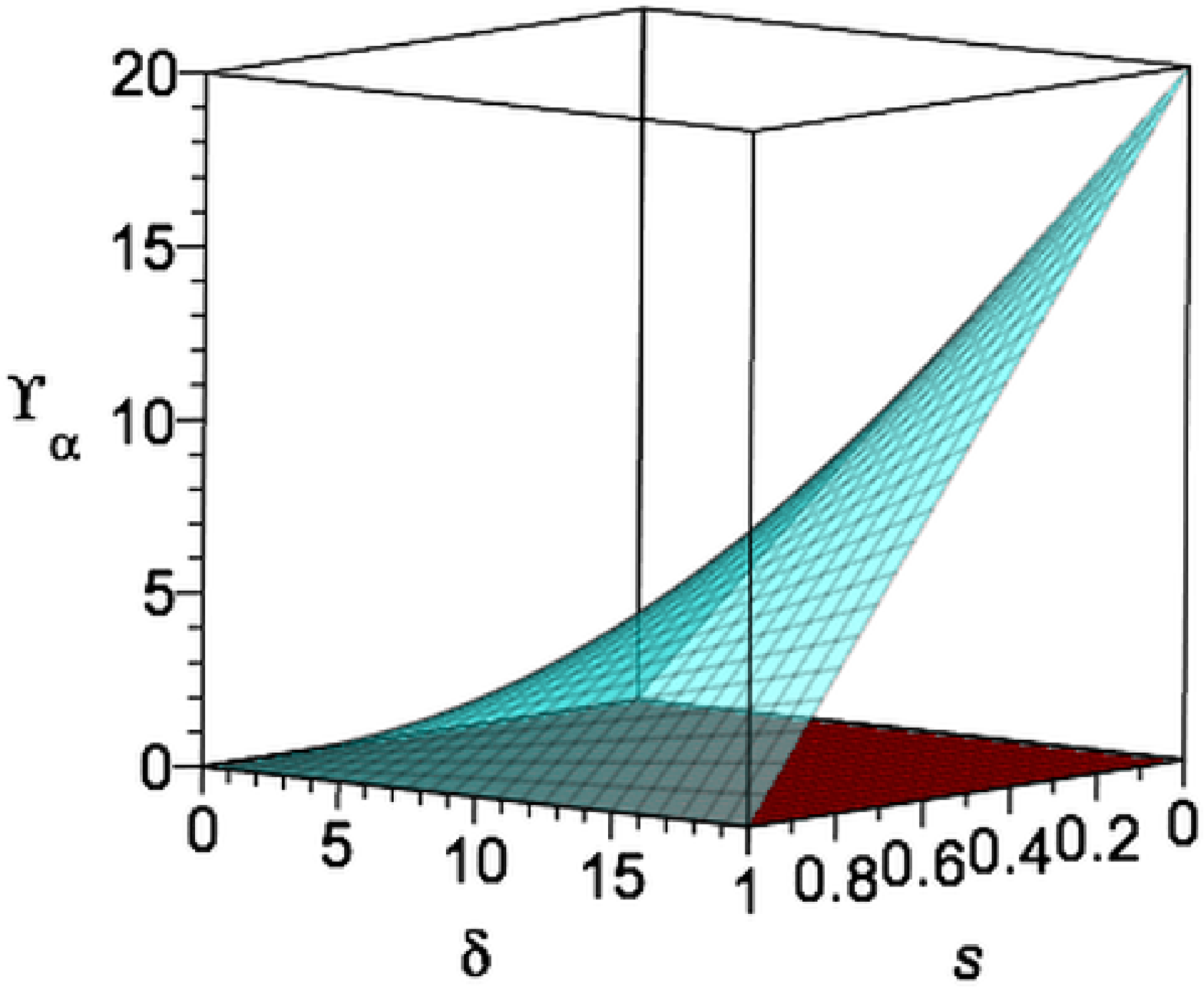}}
(b)
\caption{Width $\Upsilon_\alpha$ ($\alpha=1,2$) as the function of the decay 
rate, $\delta$,  and dimensionless time, $s=t/\tau$: $\Upsilon_1$ (red surface) 
and $\Upsilon_2$ (cyan surface). (a) $k=1$, (b) $k=512$. Parameters: $J=0.5$, 
$g=10$, $\tau = 500$, $N=1024$.}
\label{ADT3}
\end{figure}

The NQA is naturally related to the ST. Indeed, in this case the whole process of quantum annealing is described by the non-Hermitian Hamiltonian. Then, the discrete energy states of the qubit register interact with the continuum spectra (sink) of the auxillary Hamiltonian. Usually, the ST takes place in this kind of systems  when the neighboring resonances start to overlap. This means that the sum of the half-widths of the neighboring complex eigenstates is equal or exceeds the distance between these eigenstates.  In this case, the segregation of the eigenstates takes place: the eigenstate with the wide width (fast decaying) is called the ``superradiant" state, and the eigenstate with the narrow width (long-leaved) is called the ``subradient" state. The fast decay into a continuum is usually provided by the superradiant states.  The details can be found in \cite{VZA,VZ1,VZ2,Zel1,VZ3,Ber3,NABG1,BNLS,SL} (see also references therein). In application to the NQA, the situation with the ST becomes more complicated mainly because of the time-dependence of the non-Hermitian Hamiltonian.
In the NQA, one starts with the large interaction between the quantum register and the continuum (relatively large $\delta$) . This means that the initial regime of the NQA is a superradiant one. In the process of NQA, the effective interaction between the quantum register and the continuum decreases, and becomes zero at the and of the computation.
So, the system transforms from the superradiant regime (non-Hermitian) to the Hermitian one when the superradiance is absent.

To describe the ST, it is convenient to set, $\varepsilon_{\alpha} = {\mathcal E}_\alpha -i\Upsilon_{\alpha}$ ($\alpha=1,2$), where, ${\mathcal E}_\alpha= \Re \tilde E_\alpha$, and $\Upsilon_{\alpha}= -\Im \tilde E_\alpha$ is  the half-width of the resonance, $\alpha$. We denote: $\varepsilon_{1} =\varepsilon_{+}(k,t)$ and $\varepsilon_{2} =\varepsilon_{-}(k,t)$, the eigenenergies being  $\varepsilon_{\pm}(k,t)= -\varepsilon_{0k}(t) \pm \varepsilon_k(t)$. 

The results of our numerical simulations of the ST are presented in Figs.   \ref{ADT2} and \ref{ADT3}. As was mentioned above, these results demonstrate the relations between the NQA and the SR.

\section{Conclusions}

The approach presented in this 
paper, is related to application of  NQA to the ferromagnetic Ising 
spin chain. We have chosen an auxiliary Hamiltonian in such a way that the 
total Hamiltonian is non-Hermitian. At the end of evolution the non-Hermiticity vanishes. Then,  when the annealing is completed, the system is governed by the Hermitian Hamiltonian. The NQA  significantly reduces the 
time required to find the ground state of the system,  leading to the annealing 
time, $\tau \sim \ln N$, where $N$ is the number of spins (qubits), which is much better
than annealing time of Hermitian QA ($\tau \sim N^2$). We demonstrated that the NQA is related to the ST in this system. At the beginning of the NQA, the system is in the superadient state. At the end of the NQA, the total Hamiltonian becomes a Hermitian one, and the superradiance is absent. We would like to mention that because of the time-dependence of the total non-Hermitian Hamiltonian, a more detailed analysis of the relation between the NQA and the ST is required. 



\begin{acknowledgement}
 A.I.N. acknowledges the support from the CONACyT. 
 The work of G.P.B.  was carried out under the auspices of the National Nuclear Security Administration  of the U.S. Department of Energy at Los Alamos National Laboratory under Contract No.  DE-AC52-06NA25396.

\end{acknowledgement}

\bibliographystyle{plain}

\end{document}